
\magnification=1200
\hsize=15.8truecm
\vsize=23truecm
\baselineskip=12pt
\font\ttlfnt=cmcsc10 scaled 1200
\font\reffnt=cmcsc10
\font\bit=cmbxti10
\vglue 5truecm
\noindent {\ttlfnt Quantum Fragmentation}
\vskip 12pt
\noindent {\bit R. Peschanski}
\vskip 12pt
\noindent Service de Physique Th\'eorique, D.S.M.,
CEA-Saclay, 91191 Gif-sur-Yvette Cedex, France
\vskip 24pt
\noindent {\bf Abstract.} Phenomenological and theoretical aspects of
fragmentation for
elementary particles (resp. nuclei) are discussed. It is shown that some
concepts of classical fragmentation remain relevant in a microscopic
framework,
exhibiting non-trivial properties of quantum relativistic field theory (resp.
lattice percolation).
\vskip 24pt
\noindent {\bf Introduction}
\vskip 12pt
\noindent At first sight, fragmentation appears to be a typical classical
process $\lbrack$1$\rbrack$, by contrast with a quantum (or microscopic) one.
You take a \lq\lq big\rq\rq\ object of size $A,$ you break it into pieces of
different
sizes and -- if you are a scientist -- you count the mean number $ N_j $ of
pieces
of size $ j $ after a certain number of events. In rather simple but general
cases $\lbrack$2$\rbrack$, one may introduce a time-dependent description in
terms of
successive binary splittings  which can be formulated
in terms of a \lq\lq gain-loss\rq\rq\ equation, namely:
$$ { {\rm d} N_j \over {\rm d} t} = \sum^ A_{k=j+1}{\cal W}_{jk}N_k - C_jN_j\
, \eqno (1) $$
where $ {\cal W}_{jk} $ is the binary fragmentation weight
of an intermediate fragment of size $k$ into $j$ and $k-j$ and,
$$ C_j \equiv  {1 \over 2} \sum^{ j-1}_{\ell =1}{\cal W}_{\ell j} \eqno (2) $$
represents the total loss rate for the fragment species of size $ j. $
Indeed, there exist well-known papers $\lbrack$3$\rbrack$ which discuss
various mathematical
solutions of equations of the type (1-2), with applications to
e.g. depolymerisation, breaking of clusters etc...

For fragmenting quantum
objects, such as elementary particles or nuclei, however, classical
fragmentation concepts and
equations are not necessarily relevant. In these cases, the fragmenting \lq\lq
big object\rq\rq\ and the
parameter $ A $ correspond to a particle jet and its total
energy before its fragmentation into pions and other particles. For nuclei, it
is
an excited nucleus of atomic mass $A$ fragmenting into smaller ones (including
individual
 nucleons or $\alpha$ particles etc...).
Note that in the former case, one has to replace the summation by an integral
in
equations (1-2), as also considered in Refs.$\lbrack$3$\rbrack$.

There are some basic obstacles in front of us if we want to discuss
quantum fragmentation in terms of classical concepts. For
instance, for elementary particles and nuclei, the excited quantum state
which characterizes the system before its fragmentation is governed by the
same interactions (called {\sl virtual\/}) than those (called {\sl real\/})
responsible for its
subsequent
 fragmentation. It is thus by no means obvious that the classical
fragmentation
structure could emerge from the quantum environment of the
process. For instance, in field theory, the quantum fluctuations
(e.g. loop Feynman diagrams) and more generally the renormalization procedure
(necessary to give a realistic meaning to the perturbative calculations) do
not necessarily lead to an equation like (1). Quite unexpectedly, equation (1)
is
however useful for Quantum Chromodynamics (QCD), the theory of the fundamental
strong interactions between quarks and gluons, as we shall see further on.

Our aim in this
contribution is to give a brief survey of how one can cast a bridge between
quantum and classical concepts of fragmentation in this context.
We will briefly analyze two
specific cases  where the phenomenological analysis can be supported by
a theoretical model. In section 1, {\sl Quark jet fragmentation\/}
will be analyzed in the framework of
Quantum
Chromodynamics (QCD). In section 2, {\sl Nuclear multifragmentation\/}
will be  discussed in the framework
of
3-dimensional lattice percolation.
\vskip 24pt
\noindent {\bf 1. Quark Jet Fragmentation}
\vskip 12pt
\noindent The best occasion where we can observe and measure quark jet
fragmentation is
the $ {\rm e}^+ {\rm e}^- $ annihilation into hadrons (mainly pions and their
decay products)
at high energy when an intermediate $ Z^0 $ boson is formed, thereafter
decaying
into quark-antiquark $ \left(q\bar q \right) $ pairs. Such experiments have
been performed  at the LEP accelerator at
CERN.

During such reactions, in a first step lasting less than $ 10^{-24} $ second,
an
intermediary $ Z_0 $ boson is formed and decays into a $ q\bar q $ pair, often
followed by
the subsequent formation of a third (gluon) jet. These jets form
many gluons and other $ q\bar q $ pairs. This stage is well described by QCD
calculations
with a  small effective coupling
constant$\lbrack$4$\rbrack$
and can thus be studied in a quantitative
theoretical framework. Then, in a later stage of the fragmentation
process, quarks and gluons recombine into hadrons in an unknown way, only
described by modelization. The deep fundamental reason of this is that $
\alpha_ S ,$ the
effective coupling constant of QCD, happens to be time-dependent
as a
consequence of quantum fluctuations. One has
$$ \alpha_ S = {1 \over b} \left[ {\rm log}  {Q \over \Lambda_{ {\rm QCD}}}
\right]^{-1}\ , \eqno (3) $$
where $ b, $ $ \Lambda_{ {\rm QCD}}$ are fundamental constants and $ Q \sim
1/ {\rm Time} . $ One says that the
coupling constant is \lq\lq running\rq\rq\ in this theory, beeing small at
short times
and becoming large later on. One also speaks of \lq\lq
asymptotic
freedom\rq\rq\ when time is short and \lq\lq infrared slavery\rq\rq\ at long
times, since the
elementary quanta, $ q,\bar q,g, $ (called {\it partons}) are
quasi free at short times and become tightly bound at long times and
confined into
hadrons. In this limit, the theory is in a non-perturbative regime and its
complete solution is not yet known.

At short times, the predominance of quantum fluctuations and virtual
interaction effects makes difficult a classical fragmentation picture of a
quark jet. However such a description emerges from the calculations after
using a set of non-trivial properties of QCD, the field theory of Gauge
Fields describing strong interactions between partons. This theory possesses
  an internal symmetry group, local in space-time, which
is the \lq\lq color\rq\rq\ group $ {\rm SU}_3.$ We have no place for giving
the full derivation of the gain-loss equation associated with QCD, but let us
only describe it.
The main issue lies in
a system of equations which takes the form of a continuous \lq\lq
gain-loss\rq\rq\
expression similar to (1,2).

One writes:
$$ { {\rm d} D^B_A(z) \over {\rm d} \xi}  = \sum^{ }_ C \int^ 1_0 {{\rm d} x
\over x}\
P^C_A(x) \left\{ D^B_C(z/x)- x^2 \delta ^C_AD^B_A(z) \right\} \ , \eqno (4) $$
where $ D^B_A $ is the probability distribution of finding a \lq\lq
quantum\rq\rq\ $ B $ or {\sl parton\/} $ (g, $
$ q $ or $ \bar q) $ in the fragmentation of the initial \lq\lq quantum\rq\rq\
$ A $ $ (g, $ $ q $ or $ \bar q), $ with the
fraction $z$ of its total momentum. Note that Eq. (4) can be obtained,
after some manipulation, from
a continuum limit of (1) by choosing:
$$
{1 \over j} N_j = D(j/A) ;\quad {\cal W}_{jk} = {1 \over k} P(j/k);\quad t=\xi.
$$
Following the analogy with the classical process obeying Equations (1),(2),
one may
interpret the first term in the integral as a \lq\lq gain\rq\rq\ term where
the parton $ B $
is obtained via first fragmentation of $A$ into an intermediate parton $ C.$
Standard QCD
calculations give a specific prediction for the weights $ P^C_A(x) $ and thus
for
the solution of equation (4). Note that the theory also leads to a precise
re-definition of the \lq\lq evolution\rq\rq\ variable $ \xi , $ namely
$$ \xi  \equiv  \int^{ Q_{ {\rm max}}}_Q{\alpha_ S(Q) \over 2\pi} { {\rm d} Q
\over Q} = {1 \over 2\pi b} {\rm \ell n}  { {\rm \ell n}
\left( {Q_{ {\rm max}}/
\Lambda_{ {\rm QCD}}}\right) \over {\rm \ell n}  \left( {Q / \Lambda_{ {\rm
QCD}}}
\right)}.
\eqno (5) $$

{}From various experimental analyses it has been shown that equation (4) gives
a good description of the energy spectrum of jet fragmentation. However a
model-dependent piece of information has to be added since one measures
hadrons and not partons in the final state. In other physical configurations
(structure functions instead of fragmentation) the same equation holds and
can be tested with great success $\lbrack$4$\rbrack$.

Note the interesting property of equation (4) that it can be exactly solved
by the method of moments (or Mellin-transform). Introducing the moments $
\left[M_q \right]^B_A\equiv \int^ 1_0 {\rm d} z\ z^q\ D^B_A(z), $
one gets in matrix form:
$$ \left[M_q \right]\equiv {\rm exp} \left\{ \xi \left[H_q \right] \right\} \
, \eqno (6) $$
where the matrix elements $ \left[H_q \right]^B_A $ are the $q-$moments of the
weights $ P^B_A. $

The emergence of a tree structure in jet fragmentation is not only based on
the energy spectrum given by equation (4). Many other observables lead to
the same structure, while the detailed analyses show that it always implies a
non-trivial property of both the quantum and group invariance properties of
the theory. As an illustration, it was recently shown $\lbrack$5$\rbrack$ that
the
multiplicity fluctuations associated with jet fragmentation possess a
dynamical structure, similar to the intermittency phenomena in hydrodynamics,
which was predicted in particle physics some time ago $\lbrack$6$\rbrack$.
This brings an interesting analogy between quantum fragmentation
and intermittent fragmentation models, which appear in various
classical or non classical systems such as spin-glasses, polymer diffusion,
multi-particle production $\lbrack$7$\rbrack$.

Note, however, that \lq\lq differential\rq\rq$\,$  fragmentation observables
beyond the
energy spectrum and the functions $ D^B_A(z) $ are more dependent on the
unknown
\lq\lq hadronization\rq\rq\ phase of partons and thus at present more
model-dependent.  The research is going on in this field.
\vskip 24pt
\noindent {\bf 2. Nucleus Multifragmentation}
\vskip 12pt
\noindent The physical understanding of nuclear multifragmetation
is much less advanced than in QCD jet fragmentation. {\it Experimentally}
, it is only recently that systematic data on the decay products of fragmented
nuclei hence become available thanks to 4$\pi-$detectors at nuclear
accelerators
$\lbrack$8$\rbrack$. Even then, the difficulty remains of specifying without
ambiguity the excited system which multifragments, and separating its fragments
 from the pre-equilibrium particles. {\it Theoretically}, the
kinematical conditions of nuclear multifragmentation, e.g. the incident
energy and the multicomponents of the final state, are far from a known
regime of nuclear forces. One has thus to rely on models,
which are useful for the experimental investigation and may lead to
a deeper understanding of the multifragmentation phenomenon.

Among the proposed models, let us choose and discuss the one based on
3-dimensional percolation on a finite lattice$\lbrack$9,10$\rbrack$.
Due to its particular simplicity, though unexplained on a purely
nuclear-theoretical framework, it will allow us to develop on the
links between a microscopic description and classical fragmentation concepts.
This part of the talk comes from a recent study  done in collaboration
with Bertrand Giraud in Saclay$\lbrack$11$\rbrack$.

In the $3-$d percolation model, the excited nuclear system of mass $A$
(in nucleon mass units) is modelized by a finite lattice of volume $A$
with nucleon on sites and nearest-neighbour bonds. Multifragmentation results
from a breaking of these bonds
in proportion of the energy release in the system
by the reaction. Event-by-event some bonds are
randomly preserved, corresponding in
average to a ratio  $p, (0<p<1)$ of all bonds,
while a certain number of fragments are formed, giving rise
to a statistical distribution of fragments as a function of their size $i, (1
\le i < A).$ This distribution is in good agreement with recent data, if
one replaces the unknown parameter $p$ by an observable input for each event,
e.g. the multiplicity of fragments. Another parameter is introduced
corresponding to a site occupation probability, but we will stick to the
original model for simplicity.

Our aim$\lbrack$11,12$\rbrack$ is to answer the following question: is it
possible to find a classical time-dependent description of multifragmentation
which would give, at least within some approximation, the same
prediction than $3-$d
percolation concerning the distribution of fragments? In some sense,
we are looking for an eventual restoration of the time variable
in the percolation problem where such a reference
scale is absent.  Moreover, the question
behind this
is whether percolation could be described by a linear set of equations
similar to Eqns. (1-2). If such is the case, one could look for
new scaling laws, in much the same way as in the case of jet
fragmentation where they correspond to scaling properties of  QCD.

Technically, our work$\lbrack$11$\rbrack$ starts with the quest of a general,
albeit approximate, solution of the gain-loss equations (1-2). The idea
is to exhibit properties which would be independent of the particular choice
of weights ${\cal W}_{jk}$, (which we do not yet know for percolation). Then,
one looks for the same properties in some range of the percolation model to
test an eventual
 compatibility. Our conclusion is that indeed
this compatibility can be achieved during the short time evolution of the
system. Let us sketch how this can be proven.

As any such linear system, the solution of Eqns (1-2), is
otained via exponentiation once
the eigenvalues and eigenvectors of the matrix $[{\cal W}_{jk}
-{\cal C}_j \delta_{jk}]$ are determined. Note   that the matrix is triangular,
and
thus  the elements ${\cal C}
_{j}$ on the diagonale are the exact eigenvalues. However the eigenvectors
are unknown, except that they form also a triangular matrix. In fact, we were
able to prove that these vectors take the quite general form of {\it
eigenmoments}, namely they are of the form $M_{q(j)},$ where $M_q$ is the
moment of rank $q$ of the
mean distribution of fragments and $q(j) \ge 1, j=1,...,A$ are
particular, but not necessarily integer, values of the rank. One has:
$$
q(1) \equiv 1 < q(2) < q(3) ...< q(A).
$$
This property is obtained by inspection of large matrices for which the
problem is very similar to the QCD case (see Eq. (4-6)) where the moments
$M_q$
 give an exact  diagonalization in the space of particle momentum. What was
verified also for matrices of limited size, is that the
diagonalization by moments remains true for a discretized set of values
of the rank, a set ${q(j)}.$ Note however the model dependence of the set
$q(j)$ except for $q(1) \equiv 1$
which is dictated by mass conservation. A limitation of the method was
found in the case of the so-called "shattering transition",
see$\lbrack$3$\rbrack$, which needs a special treatment. With these limitations
in mind, the eigenmoment property is general enough to
be tested, e.g. in the case of percolation.

For this purpose, we remark that, if they are identified
as eigenmoments, the $M_q$'s
are linked by linear relations in Log-Log plots, and their
explicit time dependence disappears. We are thus
led to display in the same way the moments obtained
from the percolation model, choosing for instance $M_2$
for reference, see the figure. Different moments are displayed
(with $q= 1,1.5,2,3,4,5$) and show the interesting
feature of a quasi-linear dependence for the values $q=3,4,5,$ given the
fact that for $q=1$ (mass conservation) and $q=2$ (reference scale) the linear
relation is trivial. It is clear from this figure that a quasi-linear form
is obtained between $p=1$ and $p=p_c,$ where $p_c$ is the critical value above
which, in the continuous limit, an {\it infinite} percolation cluster is
formed. Indeed, the figure shows
the dominant contribution of the cluster of largest
mass to the averaged moments.
This largest cluster  is, for finite size problems,
the representative of the infinite
cluster when $p \ge p_c.$

Notice that the moments implied by the rate equations
are the {\it full} moments, including the largest fragment, while in usual
analyses of percolation models$\lbrack$9,10$\rbrack$, scaling properties are
investigated with
moments modified by the subtraction of the largest cluster.
Moreover, in such traditional analyses of percolation, the reference time
scale is generally given by the moment $M_0$ or a similar variable
related to the multiplicity of fragments. The comparison and
compatibility of our approach with such analyses is an open
problem of some interest.
\vskip 24pt
\noindent {\bf Conclusion and prospects}
\vskip 12pt
 The problem we want to discuss in conclusion of the  study of the particle and
nuclei fragmentation
is whether classical concepts of fragmentation could serve as an unifying
phenomenological picture at the microscopical level where quantum states and
fields are
involved. For this sake, let us discuss the striking common features and
differences between the two examples we have treated.

In the case of elementary particles and field theory, it is known that
only the short time evolution of a jet is accessible to perturbative
calculations. More precisely, it is only the time derivative of the
fragmentation functions (or momentum distributions of partons) which
can be calculated exactly at first order in the quantum loop expansion.
At any given fixed time, however, the knowledge of these functions
depend on an expansion at all orders (for which the renormalization
group properties can be invoked$\lbrack$4$\rbrack$). At long times
it involves the
unknown transformation of partons into hadrons. In other terms, the change in
the vacuum structure (from
partons to hadrons) prevents one from a complete theoretical understanding of
the fragmentation process.

In the case of nuclear multifragmentation, one does not possess a
comparable theoretical framework. However, the indications from
the $3-$d percolation model shows that there could be a similar property
at short times, namely the possibility of a multilinear evolution of
fragmentation. This time range corresponds to the situation when  the
fragmenting force is mild
enough to preserve the existence of at least one large cluster, that is
when $p \le p_c.$
What seems to be remarkable is the complementarity of percolation with the
previous case.
Percolation is well determined
in a given region of the parameter, namely near the
critical percolation value $p_c.$ In other terms,
fragmentation is better determined at a given "time", while
its "time-dependence", e.g. the relation between $p-$ and time-evolution
is not trivial. This is just the opposite of the field theoretical
case. It is tempting to confront the methods used in
the two cases in such a way that the stated complementarity could
hopefully be used to explore the shadow regions of both processes.

In conclusion, if fragmentation concepts could acquire some kind of
universality,
one can hope to find new methods to overcome the difficulties of
the physical description of fragmentation in the quantum world.
Some property of nuclear multifragmentation could be useful
for the hadronization problem of parton jets, as well as
perturbative methods of Quantum Field Theory could help solving
some nuclear fragmentation puzzles. More work in these directions is
deserved.

{\it Acknowledgments } Thanks are due to the organizers of the Workshop on
Fragmentation for the remarkable atmosphere they provided for discussions
and exchanges between participants coming from very different fields.
\vskip 24pt
\centerline{{\reffnt References}}
\vskip 12pt
\item{$\lbrack$1$\rbrack$}cf. These Proceedings, where many examples of
classical fragmentation studies
are discussed.
\item{$\lbrack$2$\rbrack$}See in these Proceedings, R. Botet's  talk
for a
discussion of fragmentation gain-loss equations and various generalizations.
\item{$\lbrack$3$\rbrack$}
E.D. Mc Grady and Robert M. Ziff, {\sl
Phys. Rev. Lett.\/} {\bf 58} (1987) 892,
\item{\nobreak\ \nobreak\ \nobreak\ }Z. Cheng and S. Redner, {\sl J. Phys.
A\/}: {\sl Math. Gen.\/} {\bf 23} (1990) 1233.
\item{$\lbrack$4$\rbrack$}G. Altarelli and G. Parisi, {\sl Nucl. Phys.\/}
{\bf 126} (1977) 297.
\item{\nobreak\ \nobreak\ \nobreak\ }For a complete review on
 perturbative QCD and a list of
references, {\sl Basics of
perturbative QCD\/}, by Y.L. Dokshitzer, V.A. Khoze, A.H. Mueller and S.I.
Troyan (s. Tran Than Van editors, Editions Fronti\`eres, France, 1991).
\item{$\lbrack$5$\rbrack$}Y. Dokshitzer and I. Dremin, Lund preprint
LU TP 92-30;
\item{\nobreak\ \nobreak\ \nobreak\ }W. Ochs and J. Wosiek, Nucl. Phys. {\bf
B304} (1993) 144;
\item{\nobreak\ \nobreak\ \nobreak\ }Ph. Brax, J.L. Meunier and R. Peschanski,
Nice-Saclay preprint INLN 93/1, SPhT/93/011.
\item{$\lbrack$6$\rbrack$}A. Bialas and R. Peschanski, {\sl Nucl. Phys.\/}
{\bf B273} (1986) 703.

\item{$\lbrack$7$\rbrack$}See e.g. B. Derrida and H. Spohn,
{\sl J. Stat. Phys.\/}
{\bf 51} (1988) 817 and references therein;
\item{\nobreak\ \nobreak\ \nobreak\ }J.L. Meunier and R. Peschanski, {\sl
Nucl. Phys.\/} {\bf B374} (1992) 327.
\item{$\lbrack$8$\rbrack$}See Eric Plagol's contribution, these Proceedings.
\item{$\lbrack$9$\rbrack$}
 X. Campi, {\it Phys. Lett. B}
{\bf 208} (1988) 351, and contributions to
the Proceedings of Varenna 1990 and 1992
Summer Courses of the International School of Physics {\it Enrico Fermi}.
\item{$\lbrack$10$\rbrack$}
 For a general review on percolation:
D. Stauffer, {\it Introduction to Percolation Theory}
(Taylor and Francis, London and Philadelphia, Penn. 1985.)
\item{$\lbrack$11$\rbrack$} B.G. Giraud, R. Peschanski,
{\it Eigenmoments} {\it for multifragmentation},

 Saclay T93/056 (1993), accepted for publication , {\it Phys. Lett. B}.
\item{$\lbrack$12$\rbrack$}  On a more phenomenological ground, the problem has
already  been
raised by:
\item{\nobreak\ \nobreak\ \nobreak\ } J. Richert and P. Wagner, {\it Nucl.
Phys. A}{\bf 517} (1990) 299.

\vfill
\eject
\topinsert\vskip 15.3 cm
\centerline { {\bf Percolation analyzed with the $M_2$ time scale} }
\medskip

\noindent
Relative strengths of moments
$M_q, \  q=0,1,1.5,2,3,4,5,$ as functions of $M_2$
in a Log-Log plot. Data taken from 3-d bond percolation on a $6*6*6$ lattice.
The corresponding values of the bond survival probability $p$ are shown on
the horizontal axis. Its critical value is $p_c = .25.$
Full lines: moments. Dashed lines: contributions of the largest cluster.
Dashed-dotted line: the reference moment $M_2.$
Notice that a linear behaviour is approximately obtained for
$0 \le p \le p_c$ and $q = 3,4,5$ (Figure from {$\lbrack$11$\rbrack$}).
\endinsert
\bye